\documentclass[aps,prl,reprint,groupedaddress]{revtex4-1}
\usepackage{graphicx}

\begin{document}


\title{ Late time afterglow observations reveal a collimated
  relativistic jet in the ejecta of the binary neutron star merger
  GW170817}


\author{Davide Lazzati$^1$, Rosalba Perna$^2$, Brian J. Morsony$^{3,4}$, Diego Lopez-Camara$^5$, 
Matteo Cantiello$^{6,7}$, Riccardo Ciolfi$^{8,9}$, Bruno Giacomazzo$^{9,10}$, Jared C. Workman$^{11}$}
\affiliation{ ${}^1$ Department of Physics, Oregon State University, 301 Weniger Hall, Corvallis, OR,   97331, USA \\
${}^2$ Department of Physics and Astronomy, Stony Brook University, Stony Brook, NY, 11794, USA\\
${}^3$ Department of Astronomy, University of Maryland, 1113 Physical Sciences Complex, College Park, MD 20742-2421, USA \\
${}^4$ AAAS Science and Technology Policy Fellow \\
${}^5$ CONACYT---Instituto de Astronom\'ia, Universidad Nacional Aut\'onoma de M\'exico, A.P. 70-264, 04510 M\'exico D.F., M\'exico\\
${}^6$ Center for Computational Astrophysics, Flatiron Institute, 162 5th Avenue, New York, NY 10010, USA\\
${}^7$ Dept. of Astrophysical Sciences, Peyton Hall, Princeton University, Princeton, NJ 08544, USA\\
${}^8$ INAF, Osservatorio Astronomico di Padova, Vicolo dell'Osservatorio 5, I-35122 Padova, Italy\\
${}^9$ INFN--TIFPA, Trento Institute for Fundamental Physics and Applications, Via Sommarive 14, I-38123 Trento, Italy\\
${}^{10}$ Physics Department, University of Trento, via Sommarive 14, I-38123 Trento, Italy \\
${}^{11}$  Department of Physical and Environmental Sciences, Colorado Mesa University, Grand Junction, CO 81501, USA
}

\date{\today}

\begin{abstract} 
  The binary neutron star (BNS) merger GW170817 was the
  first astrophysical source detected in gravitational waves and
  multi-wavelength electromagnetic radiation. The almost simultaneous
  observation of a pulse of gamma-rays proved that BNS mergers are
  associated with at least some short gamma-ray bursts (GRBs).
  However, the gamma-ray pulse was faint, casting doubts on the
  association of BNS mergers with the luminous, highly relativistic
  outflows of canonical short GRBs. Here we show that structured
  jets with a relativistic, energetic core surrounded by slower and
  less energetic wings produce afterglow emission that brightens
  characteristically with time, as recently seen in the afterglow of
  GW170817. Initially, we only see the relatively slow material moving
  towards us. As time passes, larger and larger sections of the
  outflow become visible, increasing the luminosity of the afterglow.
  The late appearance and increasing brightness of the
  multi-wavelength afterglow of GW170817 allow us to constrain the
  geometry of its ejecta and thus reveal the presence of an off-axis
  jet pointing about $30^\circ$ away from Earth. Our results confirm a
  single origin for BNS mergers and short GRBs: GW170817 produced a
  structured outflow with a highly relativistic core and a canonical
  short GRB. We did not see the bright burst because it was beamed
  away from Earth. However, approximately one in 20 mergers detected
  in gravitational waves will be accompanied by a bright, canonical
  short GRB. 
\end{abstract}


\maketitle

\section{Introduction}

The almost simultaneous detection of gravitational waves (GW170817)
and of gamma-rays (GRB170817A) as a result of the merger of two
neutron stars in a binary
\cite{Abbott2017a,Abbott2017b,Goldstein2017,Savchenko2017} has been
followed by a massive observational campaign covering a wide portion
of the electromagnetic spectrum
\cite{Coulter2017,Hallinan2017,Soares2017,Troja2017,Margutti2018,Mooley2018,Ruan2018}.
Early UV, optical, and IR detections were obtained within one day of
the GW trigger \cite{Coulter2017,Soares2017}. Their spectra and
temporal evolution were shown to be consistent with the quasi-thermal
radiation from a kilonova \cite{Kasen2017,Pian2017}, a transient
powered by the radioactive decay of heavy nuclei synthesized within
the merger ejecta.

In the X-rays, the source was detected 9 days after the GW event by
Chandra \cite{Troja2017}, while radio emission was detected a few days
later \cite{Hallinan2017}. X-rays and radio emission were
characterized by non-thermal spectra from a single power-law spanning
more than eight orders of magnitude in frequency. This indicated a
common origin for the high- and low-frequency emission consistent with
the afterglow from a relativistic blast-wave \cite{Meszaros1997}. The
early X-ray and radio observations were consistent with a diverse set
of models for the origin of the blast-wave, including a top-hat jet
seen off-axis
\cite{Alexander2017,Haggard2017,Ioka2017,Kim2017,Margutti2017,Murguia2017b,Troja2017,Xiao2017,He2018},
a mildly relativistic, isotropic fireball \cite{Salafia2017}, and a
structured jet \cite{Kasliwal2017,Lamb2017,Lazzati2017a}.

Continued monitoring revealed that the afterglow luminosity is
steadily increasing with time \cite{Margutti2018,Mooley2018,Ruan2018}
a behavior that is anomalous for canonical GRBs \cite{Nousek2006} and
rules out simple models like the off-axis top-hat jet and the
isotropic fireball \cite{Mooley2018} (see Results). Any viable
explanation requires the continued injection of energy in the external
shock, which can be accomplished in different ways. If the central
source left from the merger were still active (for example a
magnetar), its continued energy release would energize the external
shock \cite{Ai2018,Geng2018,Li2018}. Alternatively, radially stratified ejecta could
provide a source of energy as the slower material catches up with the
external shock either with \cite{Mooley2018} or without \cite{Hotokezaka2018} the presence
of a jet. This requires the presence of fast
ejecta from the BNS merger that are significantly stratified with a
very steep energy profile $E(\beta\gamma)\sim (\beta\gamma)^{-5}$,
with a high-speed cut-off of at least $\beta\sim 0.5$. The mechanism
that could accelerate the ejecta to such large speed is, however, not
well understood.

Structured jets, on the other hand, are a natural outcome of BNS
mergers, irrespectively of the detailed properties of the ambient
material surrounding the merger site
\cite{Aloy2005,Lazzati2017b,Gottlieb2018,Kathi2018,Mooley2018}. The
propagation of a light relativistic jet through a dense environment
drives a forward/reverse shock system that causes the production of a
cocoon around the jet. The cocoon has high pressure but no bulk
relativistic motion, shearing the jet/cocoon boundary. The ensuing
structure is characterized by a narrow, highly relativistic jet,
surrounded by a sheath of light but slower material and mildly
relativistic wings at large angles \cite{Lazzati2017b}. While
structured jets can be obtained with other mechanisms, this jet-cocoon
mechanism is quite general and is guaranteed to produce a structured
jet, fairly independent of the initial structure of the outflow and
the amount of ambient material it travels through
\cite{Aloy2005,Lazzati2017b,Gottlieb2018,Kathi2018}. After it has
released the prompt emission either at the photosphere
\cite{Lazzati2017a,Lazzati2017b} or, more likely, via shock
dissipation processes \cite{Kasliwal2017}, the structured jet
propagates and drives an external shock into the interstellar medium
\cite{Lamb2017,Jin2018}.  Non-thermal particles and magnetic fields
generated downstream the shock produce synchrotron radiation, the
broadband emission that is commonly referred to as afterglow
\cite{Meszaros1997,Sari1998}. A structured jet seen on-axis is
indistinguishable from a top-hat jet; if, instead, the jet is seen
off-axis, differences in its structure become apparent
\cite{Granot2002,Rossi2004,Lamb2017}.

Here, we compute multi-wavelength afterglow light curves from the
structured fireball expected to develop from a BNS merger
\cite{Lazzati2017b} that also produces a canonical SGRB. We then
compare the calculated light curves to the available dataset for
GW170817 and show that the model is in agreement with the data. With
the afterglow data only, we can constrain the viewing angle to within
an uncertainty of only a few degrees and we find it to be in agreement
with previous constraints from independent estimates. The prompt
  emission from the model that we adopt is discussed in
  \cite{Lazzati2017b}, where it is shown that the energetics and
  duration of the pulse are consistent with the observations of
  GW170817 \cite{Goldstein2017,Savchenko2017}. There is tension,
  instead, between the predicted frequency (a few keV, in the soft
  X-rays) and the observations ($\sim200$~keV, in the soft
  gamma-rays). The discrepancy is due to the assumption of a
  dissipationless cocoon in \cite{Lazzati2017b}. The observed
  transient, instead, calls for a shocked cocoon in which the radiation
  is released by the breakout of a radiation dominated shock
  \cite{Nakar2012,Kasliwal2017}.

\begin{figure}
\parbox{0.49\columnwidth}{
\includegraphics[width=0.48\columnwidth]{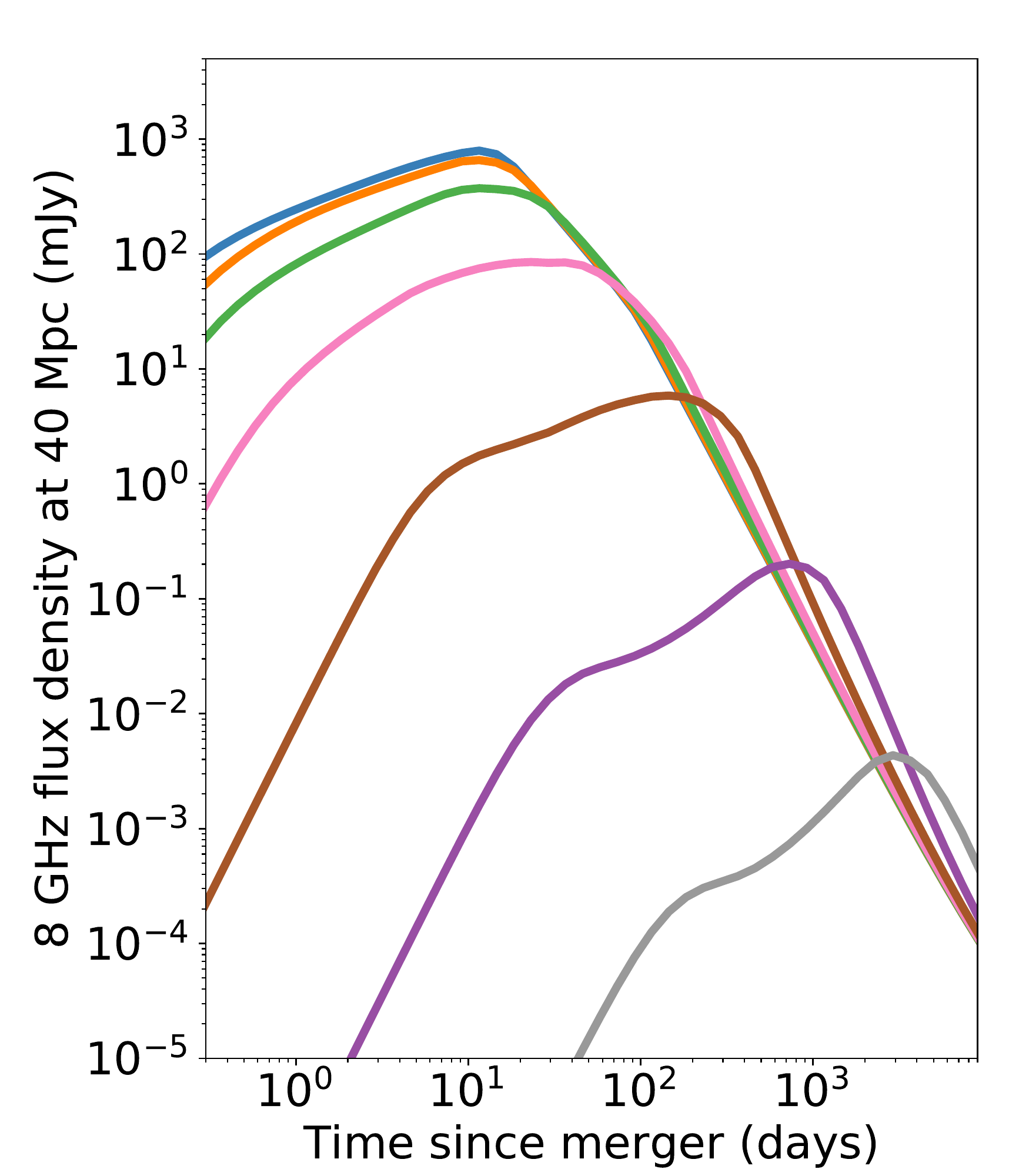}}
\parbox{0.49\columnwidth}{
\includegraphics[width=0.48\columnwidth]{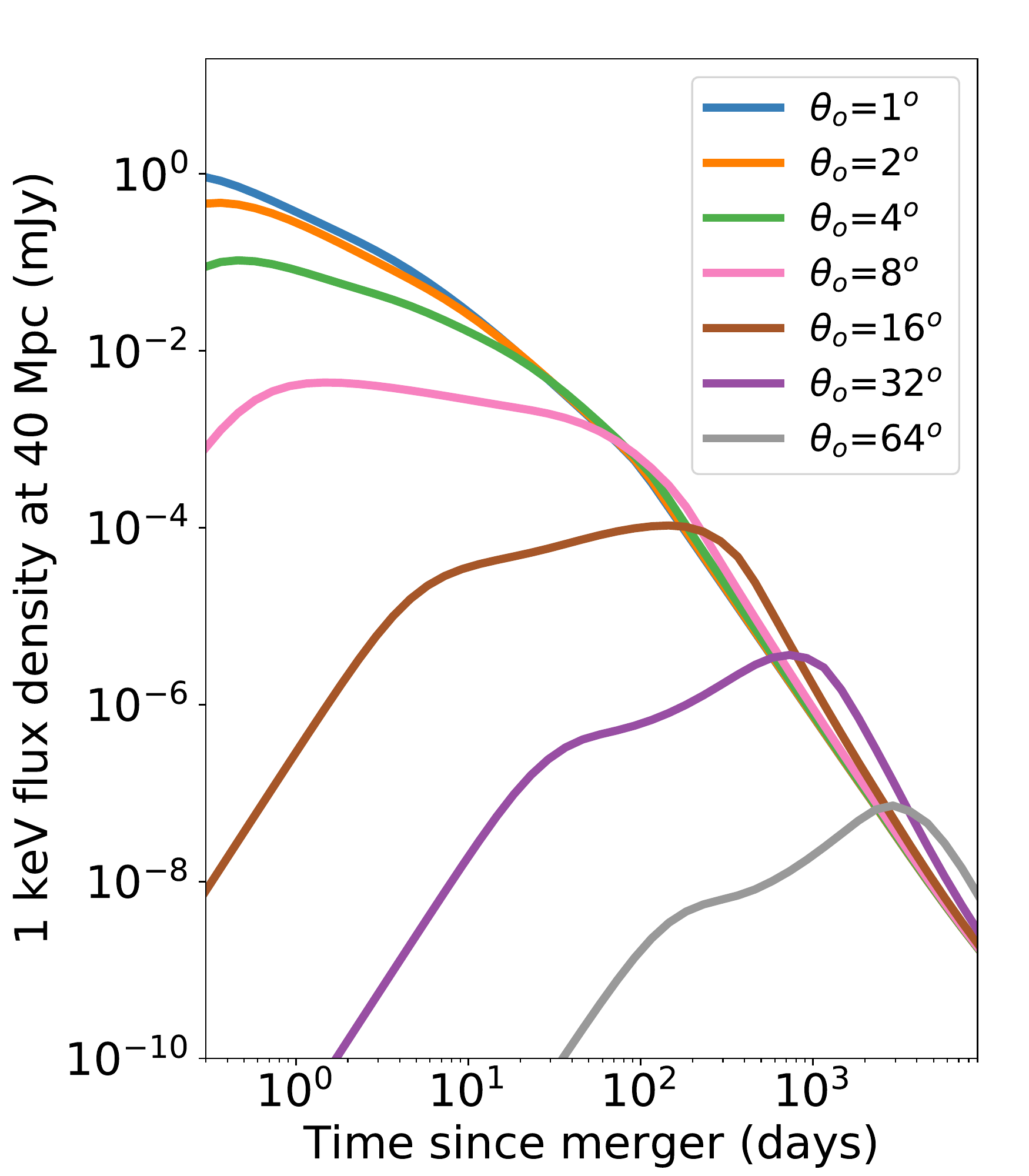}
}
\caption{{Light curves of the afterglow of a structured jet calculated
    at a frequency of 8 GHz (left panel) and 1 keV (right panel) for
    several observers. Viewing angles are indicated in the legend. The
    parameters used for these models are: $\epsilon_e=10^{-2}$,
    $\epsilon_B=10^{-3}$, $p_{el}=2.3$, and
    $n_{ISM}=10^{-4}$~cm$^{-3}$. }
 \label{fig:lcs}}
 \end{figure}

\section{Methods}
\subsection{Calculation of the afterglow light curves}

Light curves and spectra from the external shock were computed
adopting standard techniques
\cite{Sari1998,Panaitescu2000,Granot2002,Rossi2004}. Our code
integrates the emission over the equal arrival times surface
\cite{Panaitescu1998} and assumes that different sections of the jet
do not undergo sideways expansion \cite{Rhoads1997} even after the jet
comes in causal contact with its boundary (when
$\Gamma < \theta_j^{-1}$). The amount of sideways expansion is a
debated topic, with numerical simulations of top-hat jets suggesting a
limited amount of spreading, even at late times \cite{Eerten2010}. In
any case, sideways expansion has a small impact on the light curves
and the assumption that we made does not affect our conclusions (see,
e.g., Figure 4 of Rossi et al. \cite{Rossi2004}). 

The input data of our structured jet model are taken from a numerical
simulation of a jet propagating through non-relativistic ejecta from a
binary neutron star merger \cite{Lazzati2017b}.  The simulation was
carried out prior to the detection of GW170817. It was initiated in 3D
and subsequently mapped in 2D for the large scale
evolution. Cylindrical symmetry was assumed for the afterglow
calculations. The input parameters of the simulation -- jet energy
($E=10^{50}$ erg) and opening angle ($\theta_j=16^\circ$), yielding
$E_{\rm{iso}}=2.6\times10^{51}$ -- were chosen to mimic as best as
possible a typical short GRB \cite{Fong2015}. The jet leaving the
ambient material of the merger has been hydrodynamically collimated
into a narrower $theta_j=5^\circ$ cone, with an isotropic equivalent
energy $E_{\rm{iso}}=10^{52}$~erg. Previous studies
\cite{Murguia2014,Nagakura2014,Duffell2015,Murguia2017} had confirmed
that the jet can survive the interaction, at least in some cases, and
in this work we concentrate on the polar structure of the ejecta at
large radii.

\begin{figure}
\includegraphics[width=\columnwidth]{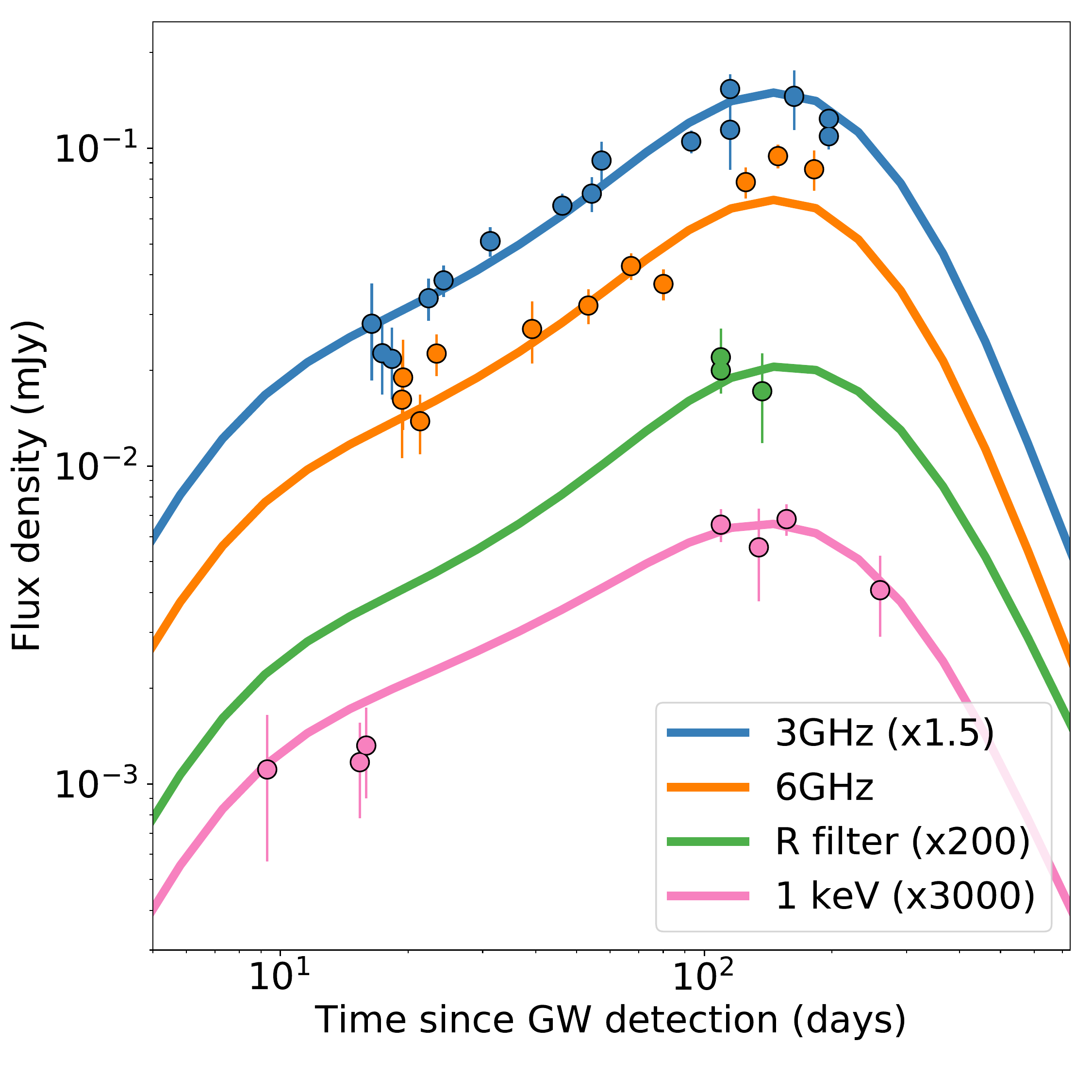}
\caption{{The radio, optical, and X-ray light curves from a structured
    jet model are shown in blue (3 GHz), yellow (6 GHz), green (R
    band) and pink (X-rays), respectively. For viewing purposes, all
    but the 6 GHz curves have been multiplied by a constant (as
    specified in the legend). Radio and X-ray measurements with
    uncertainties are shown with filled symbols.  Additional data at
    other radio frequencies have been used in the fit but are not
    shown because they are sparse and would overcrowd the figure. The
    optical light curve at early time was dominated by the kilonova
    emission, which is not considered in our modelling. The model
    shown has $\chi^2=69$ with 56 degrees of freedom. The observer
    lies at an angle of $33^\circ$ from the jet axis and the fireball
    propagates in a uniform external medium with number density
    $n_{\rm ISM}= 4.2\times10^{-3}$~cm$^{-3}$. The figure includes the
    most recent Chandra observation (\cite{Alexander2018}, the X-ray
    datum at 260 days), which is not fit but just overlaid on the best
    fit curves from the data at prvious times.}
\label{fig:bestfit}}
\end{figure}

As typical for afterglow calculations, our models depend on the
microphysical shock parameters that describe the particle distribution
and magnetic field intensity downstream the shock. A fraction
$\epsilon_e$ of the shock energy is given to electrons, and all the
electrons are accelerated in a power-law distribution
$n(\gamma)\propto \gamma^{-p_{el}}$. In addition, a fraction
$\epsilon_B$ of the shock energy is assumed to be converted into a
tangled magnetic field. Given the location of GW170817 in the outskirt
of an early type galaxy \cite{Coulter2017,Soares2017}, we compute our
models for a uniform interstellar medium of number density
$n_{\rm ISM}$. The last free parameter is the orientation of the line
of sight with respect to the jet axis. Neither the blastwave energy
nor its initial Lorentz factor are free parameters in our model. They
are both determined uniquely by the polar angle. Figure~\ref{fig:lcs}
shows the resulting radio (8 GHz) and X-ray (1 keV) light curves for
different viewing angles. The same code was used also for the
calculation of afterglow light curves from top-hat jets and isotropic
fireballs. A top-hat model is a jet with constant energy and Lorentz
factor within a specified jet angle $\theta_j$. The jet has sharp
edges, and the energy and velocity outside of the jet angle are set to
zero.

In the case of isotropic fireballs, the model has six free parameters
(one more than for structured jets) since the blastwave energy $E$ and
the initial Lorentz factor $\Gamma_0$ are free, but the observer angle
is irrelevant. In the case of a top-hat jet, the model has seven free
parameters. In addition to the five of the structured jet, one has to
consider the jet's isotropic equivalent energy $E_{\rm iso}$ and the
jet opening angle $\theta_j$. Since we are concerned with top-hat jets
seen off-axis, the initial Lorentz factor is irrelevant, and we set it
to $\Gamma_0=300$ for all top-hat models. While more sophisticated
models for top hat jets are available \cite{Eerten2011}, in order to
be able to directly compare the isotropic, structured, and top hat
jets it is fundamental to adopt the same code in the calculation of
the three.

\begin{figure*}
\parbox{0.49\textwidth}{
\includegraphics[width=0.48\textwidth]{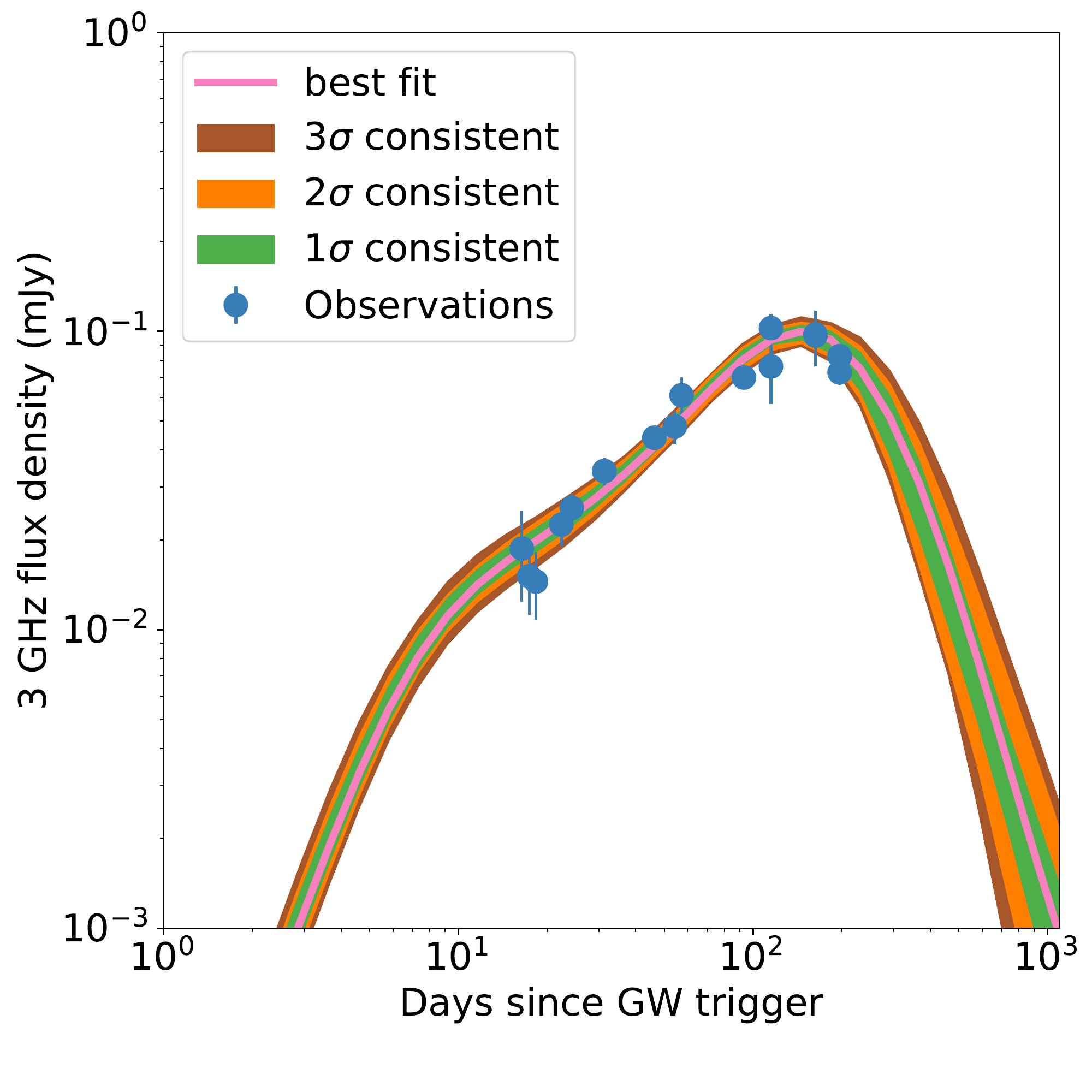}}
\parbox{0.49\textwidth}{
\includegraphics[width=0.48\textwidth]{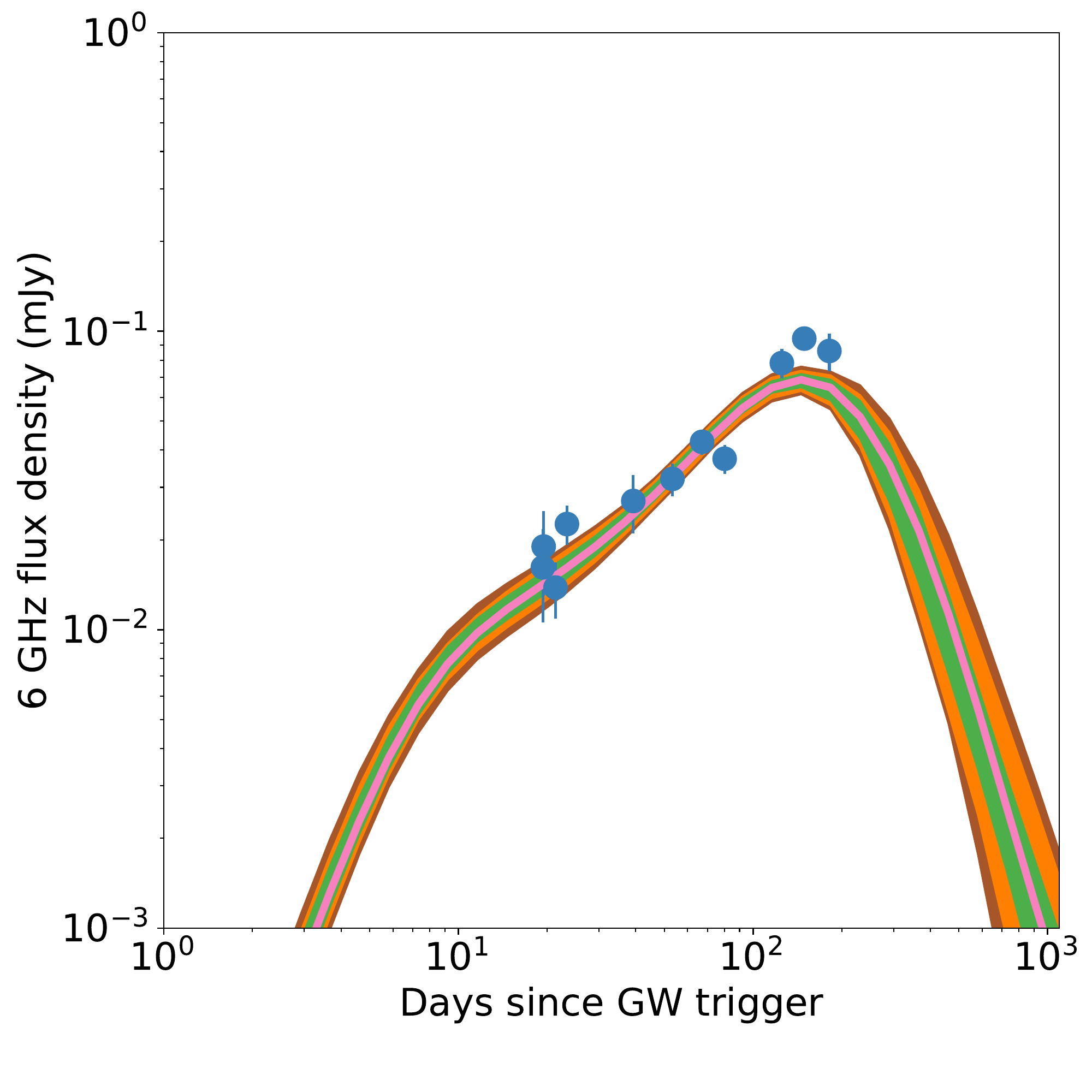}}
\parbox{0.49\textwidth}{
\includegraphics[width=0.48\textwidth]{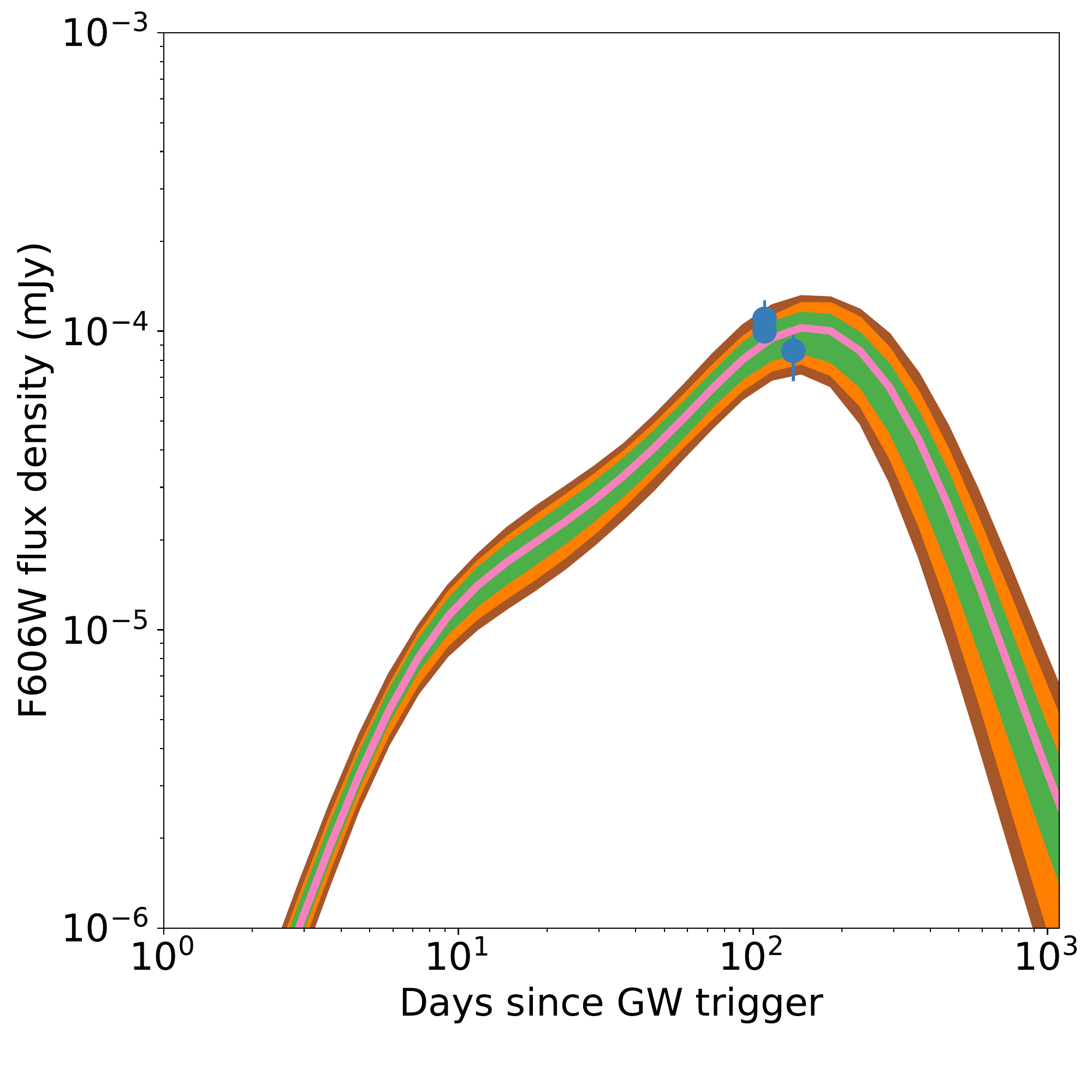}}
\parbox{0.49\textwidth}{
\includegraphics[width=0.48\textwidth]{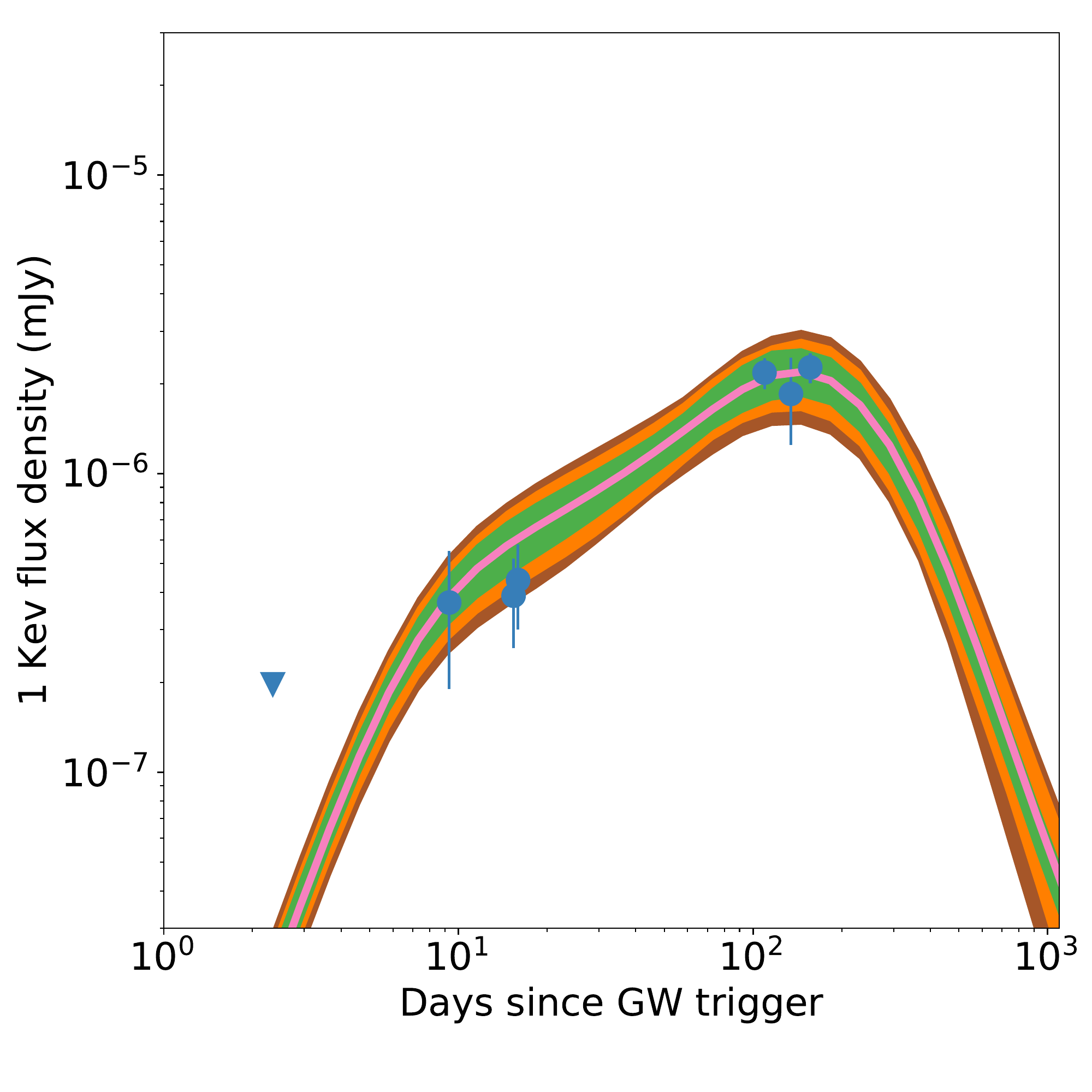}}
\caption{{Light curves and their uncertainty areas compared with the 
    measurement of the afterglow of GW170817. The pink line is the
    best fit model. The green shaded area is the envelope of all
    models consistent with the data at the $1\sigma$ level,
    orange is consistent at the $2\sigma$ level, and brown is
    consistent at the $3\sigma$ level. Data are displayed with
    solid blue markers. All models were fit simultaneously to the
    entire multi-wavelength dataset. The upper left panel shows the
    3~GHz light curve, the upper right panel shows the 6~GHz light
    curve, the bottom left panel shows the 606~nm light curve, and the
    bottom right panel shows the 1~keV light curve.} 
\label{fig:behave}} \end{figure*}

\subsection{Data modelling}

\begin{figure}
 \includegraphics[width=\columnwidth]{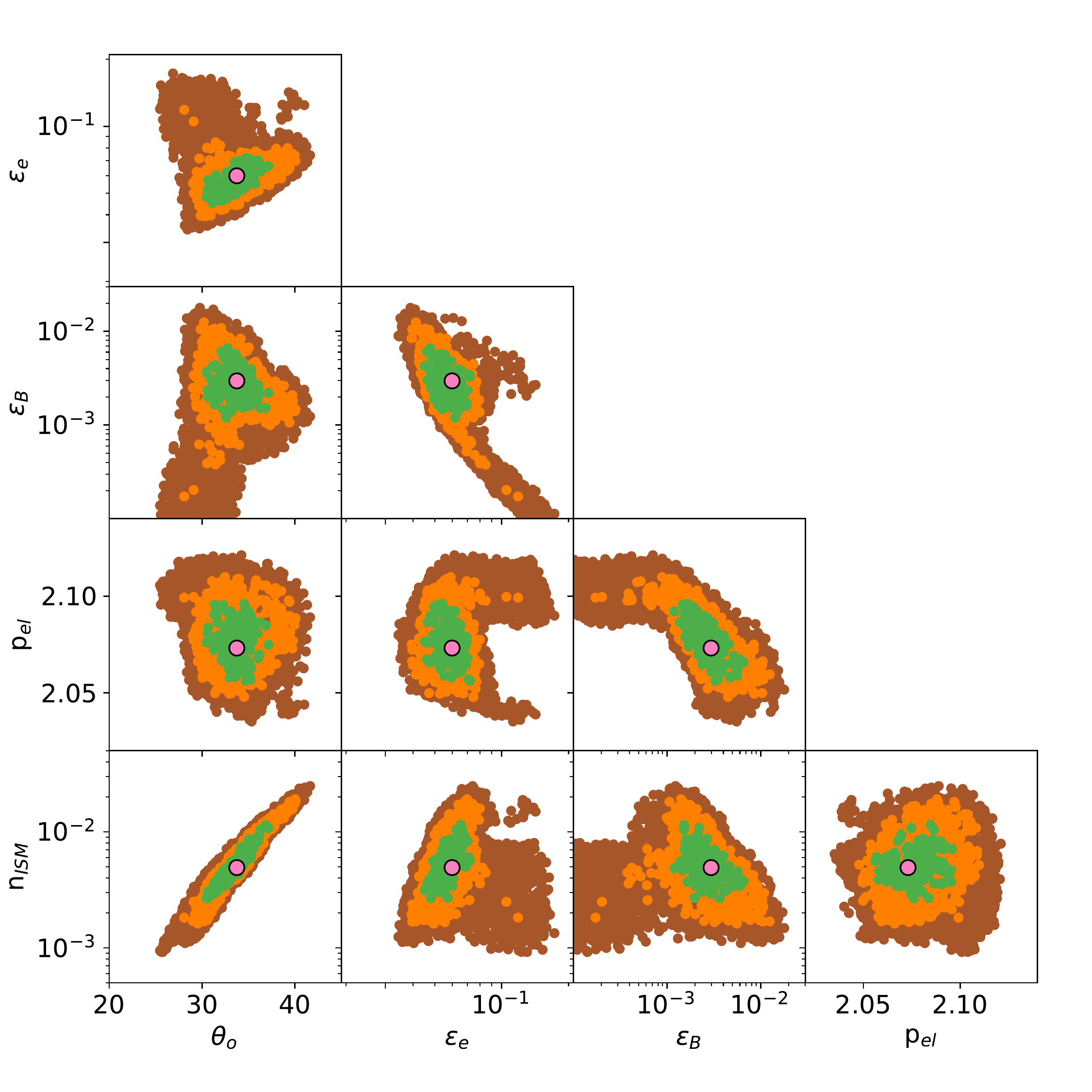}
 \caption{ Corner plot showing the degeneracy of the fit parameters
   for the structured jet model. 
   A strong degeneracy is evident between the
   interstellar density and the observing angle. The meaning of different colors is
   the same as in Figure 3.
   \label{fig:corner}}
 \end{figure}

Several previous publications have considered structured jets as a
plausible explanation for the data of GW170817
\cite{Granot2017,Xiao2017,Granot2018,Lamb2018,Zou2018}, while only one
publication presents a formal fit to the data analogous to what we
present here \cite{Resmi2018}, however their structured model is
analytical and not self-consistently derived from simulations of jet
propagation.  Our fitting dataset is the result of a complete
collection (to our best knowledge) of data published in the literature
\cite{Alexander2017,Haggard2017,Margutti2017,Hallinan2017,Troja2017,Dobie2018,Lyman2018,Margutti2018,Mooley2018,Ruan2018,Troja2018}.
Since the calculation of top-hat and structured jet models is
numerically intensive, we have performed a Markov Chain Monte Carlo
analysis. We have constructed a grid of models for all three cases for
a set of parameter values. We then draw a random selection of the
parameter set and perform a multi-linear interpolation of the model
grid in log space to derive the model at the desired value of the
parameters. We then compute the $\chi^2$ of said model with respect to
the data. The parameter set is defined to be behavioral if the
probability value is more than 0.0027, corresponding to $3\sigma$.
Figure~\ref{fig:bestfit} shows the best fit model for the structured
jet from Table~\ref{tab:bestfit} compared to data in four selected
bands: 3 GHz, 6 GHz, optical, and 1 keV. Figure~\ref{fig:behave} shows
all the behavioral models in the same bands in four panels. Models are
color-coded according to their probability. The corner plot of
parameter degeneracy is shown in Figure~\ref{fig:corner}. It shows
that there is a pronounced degeneracy between the interstellar density
and the viewing angle. Finally, in Figure~\ref{fig:comp} we compare
the best fits of the three models with one another.

\section{Results}
We have computed the afterglow light curve (see Methods and
Figure~\ref{fig:lcs} and~\ref{fig:behave}) from a structured jet
obtained from a numerical simulation of relativistic outflows from
binary NS mergers \cite{Lazzati2017b}. A frame of the simulation is
shown in the left panel of Figure~\ref{fig:decomp} and the energy and
Lorentz factor profiles are shown in the lower right panel of the same
figure. The structured jet from our simulation does not have a
well-identified core, unlike a top-hat jet. As seen in the lower panel
of Figure~\ref{fig:decomp}, the core of the jet extends out to an
off-axis angle of approximately $5^\circ$, which carries
$\sim 10^{52}$~erg of isotropic equivalent energy. These properties
are on the high side of short GRB population studies, yet consistent
with a short GRB observed on-axis \cite{Ghirlanda2016}, especially
considering that only a fraction of the kinetic energy is converted to
radiant energy. The core Lorentz factor of order 100 is also
consistent with constraints from the prompt emission of on-axis events
\cite{Nakar2007}. With the exception of the on-axis models with
$\theta_o<10^\circ$, all our synthetic light curves are characterized
by a fast early phase (proportional to $t^3$), a break at a few days
followed by a slow luminosity increase, a maximum at a few months to
years after the merger, and a final decay over several years (see
Figure~\ref{fig:lcs}). The material that travels directly along the
line of sight is responsible for the early emission \cite{Mooley2018}.
The first break at a few days is due to the deceleration of that
material once it accumulates enough interstellar mass. As time
progresses, material that travels at increasingly large angles with
respect to the line of sight decelerates and its radiation becomes
visible. To the observer, it appears that the fireball's energy has
increased and therefore the afterglow brightens. Eventually, the jet
core that carries most of the energy comes into view. This corresponds
to the maximum in the light curves and happens between a few months
and a few years after the merger, depending on a combination of the
jet's energy and Lorentz factor and of the interstellar density.
Figure~\ref{fig:decomp} shows a decomposition of the 3 GHz light curve
in five components. Most of the radiation comes from the region of the
outflow between the line of sight and the core (regions A, B, and C in
the figure), the outer regions (D and E) contributing a negligible
amount of flux.

\begin{figure}
\includegraphics[width=\columnwidth]{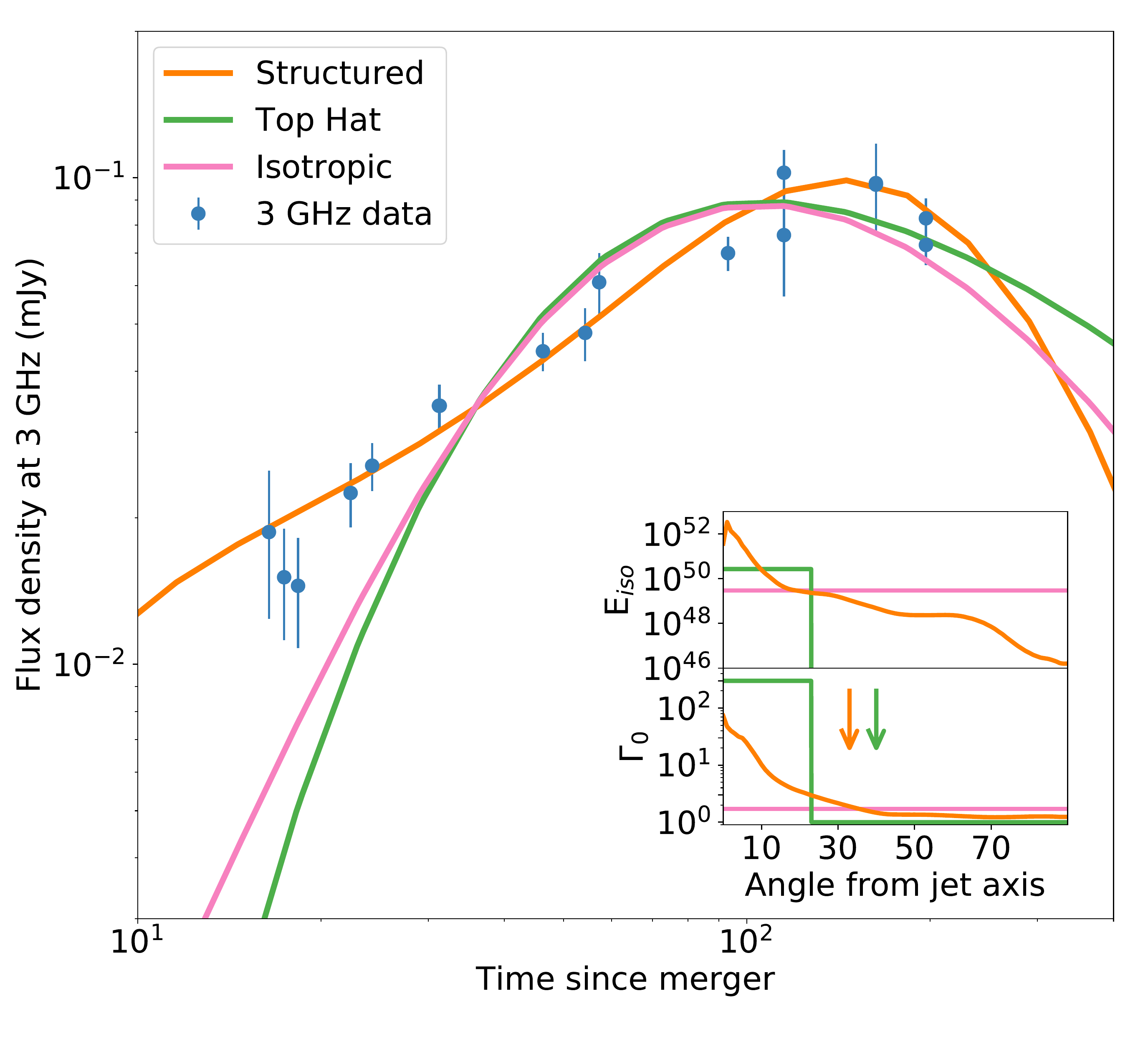}
\caption{Comparison among the best-fits 3~GHz light curves for the three model 
structures discussed: an isotropic fireball (pink), an off-axis,
top-hat jet (green), and a structured jet (orange). Data at 3 GHz 
are shown with solid blue symbols, but the fits were performed on all 
the multi-wavelength dataset. The inset shows the best-fit energy 
and Lorentz factor profiles of the three models. The vertical arrows 
show the location of the observer in the top-hat and structured models.
\label{fig:comp}}
 \end{figure}

 The radio, optical, and X-ray light curves from the best-fit model
 are shown in Figure~\ref{fig:lcs} and the spectra at two epochs are
 shown in Figure~\ref{fig:spex}. Data are shown for comparison on both
 figures. The best-fit model has a statistically acceptable $\chi^2$
 (69 for 56 degrees of freedom, with probability $p=11\%$) and it is
 characterized by a viewing angle $\theta_o=33^{+4}_{-2.5}$ in
 degrees, consistent with other constraints on the viewing geometry.

In addition, we have also attempted to fit a low-$\Gamma$ isotropic
fireball (without any radial stratification) and a top-hat jet model
to the data. We find that these two models can be rejected at very
high statistical confidence. The comparison of the best fit of the
three models is shown in Figure~\ref{fig:comp}, and the details of the
fits are reported in Table~\ref{tab:bestfit}.

\begin{figure}
 \includegraphics[width=\columnwidth]{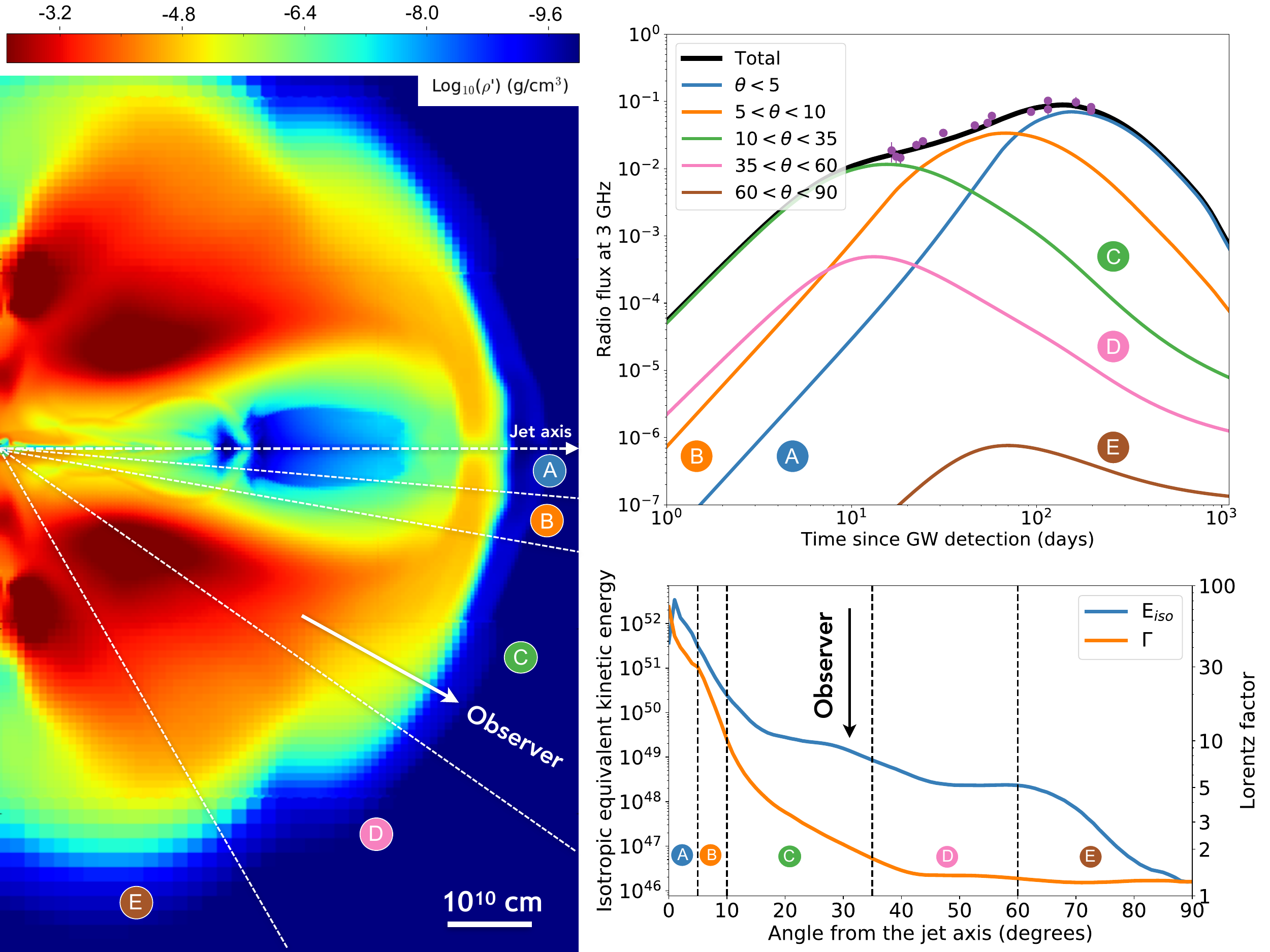}
 \caption{Left panel: pseudocolor density image of the hydrodynamic
   numerical simulation of a short gamma-ray burst jet from a binary
   neutron star merger used to compute the afterglow light curves
   \cite{Lazzati2017b}. The propagation of the jet through
   non-relativistic ejecta in the vicinity of the merger site causes
   the emergence of a structured jet, with a light, fast core (the
   low-density, blue region) and less energetic, slower wings (the
   orange and green material surrounding the jet). The line of sight
   to the observer (lying at 33$^\circ$ from the jet axis) is shown
   with a white arrow. The polar distribution of energy and velocity
   of the ejecta is shown in the bottom right panel. The top right
   panel shows the best fit afterglow model decomposed into radiation
   coming from the core of the jet (blue), the fast wings of the jet
   (orange), the material moving along the line of sight (green), and
   material at large angles (pink and brown, which do not contribute
   to the observed afterglow emission). The solid black line is the
   sum of the colored lines.
   \label{fig:decomp}}
 \end{figure}
 
\begin{table}
\begin{tabular}{l|c|c|c}
& \quad Isotropic \quad & \quad Top Hat \quad & \quad Structured \quad\\ \hline
$\chi^2$/d.o.f. & 286/55 & 266/54 & 69/56 \\
probability & $<10^{-10}$ & $<10^{-10}$ & 0.11 ($1.6\sigma$) \\
$E_{\rm{iso}}$ (erg) & $2.5\times10^{49}$ & $5.7\times10^{51}$ & -- \\
$\Gamma_0$ & 1.8 & -- & -- \\
$\theta_j$ (degrees) & -- & 23 & --\\
$\theta_o$ (degrees) & -- & 39 & $33^{+4}_{-2.5}$ \\
$\epsilon_e$ & 0.03 & 0.01 & 0.06$\pm0.01$ \\
$\epsilon_B$ & 0.0002 & 0.003 & 0.0033$\pm$0.002 \\
$p_{\rm{el}}$ & 2.065 & 2.07 & 2.07$\pm$0.01 \\
$n_{\rm{ISM}}$ (cm$^{-3}$) & $6.8\times10^{-4}$ & $9.1\times10^{-4}$ & $(4.2^{+8.5}_{-1.6})\times10^{-3}$
\end{tabular}
\caption{{Parameters and statistical properties of the best fit for
    the three models analyzed.}
\label{tab:bestfit}}
\end{table}

\section{Conclusions}

The goodness of our fit and the consistency of our results with other
independent measurements strongly suggest that binary neutron star
mergers do produce short GRBs, as predicted many years ago
\cite{Eichler1989} and supported by several lines of circumstantial
evidence from their host galaxies, locations, and rate
\cite{Berger2014}. We found that within the structured jet-cocoon
model, all observations can be reconciled if the observer lied at a
viewing angle $\theta_o\approx30^\circ$. Such viewing
geometry is consistent with the GW amplitude \cite{Abbott2017a}
($\theta_o\lesssim30^\circ$), with the prompt gamma ray energetics and
duration \cite{Lazzati2017b}
($20^\circ\lesssim \theta_o\lesssim40^\circ$), the kilonova
characteristics \cite{Perego2017} ($15^\circ\lesssim \theta_o\lesssim35^\circ$), and
the afterglow modelling presented here.

Future observations could lend further support to this conclusion.
First, due to its relative proximity to Earth and radio brightness,
the remnant of GW170817 can be resolved with the VLBI
\cite{Gill2018,Nakar2018}. We expect that at the time of maximum
afterglow luminosity the physical size of the remnant will be
approximately one parsec, or 4.6 milli arcsec, compared to the VLBI
angular resolution of 0.3 milli arcsec at 30 GHz. Asymmetry in the
ejecta brightness should be prominent before and around the peak time
in the structured jet scenario, while the remnant would be fairly
spherical in the radially stratified case. Linear polarization with
consistent position angle peaking around the maximum brightness of the
afterglow would also confirm the presence of a jet
\cite{Rossi2004,Gill2018}. Finally, direct confirmation of the
association is possible. Once every 20 events, we expect the line of
sight to be within the jet cone and a bright, on-axis short GRB to be
observed in coincidence with a GW signal from the merger.

\begin{figure}
\includegraphics[width=\columnwidth]{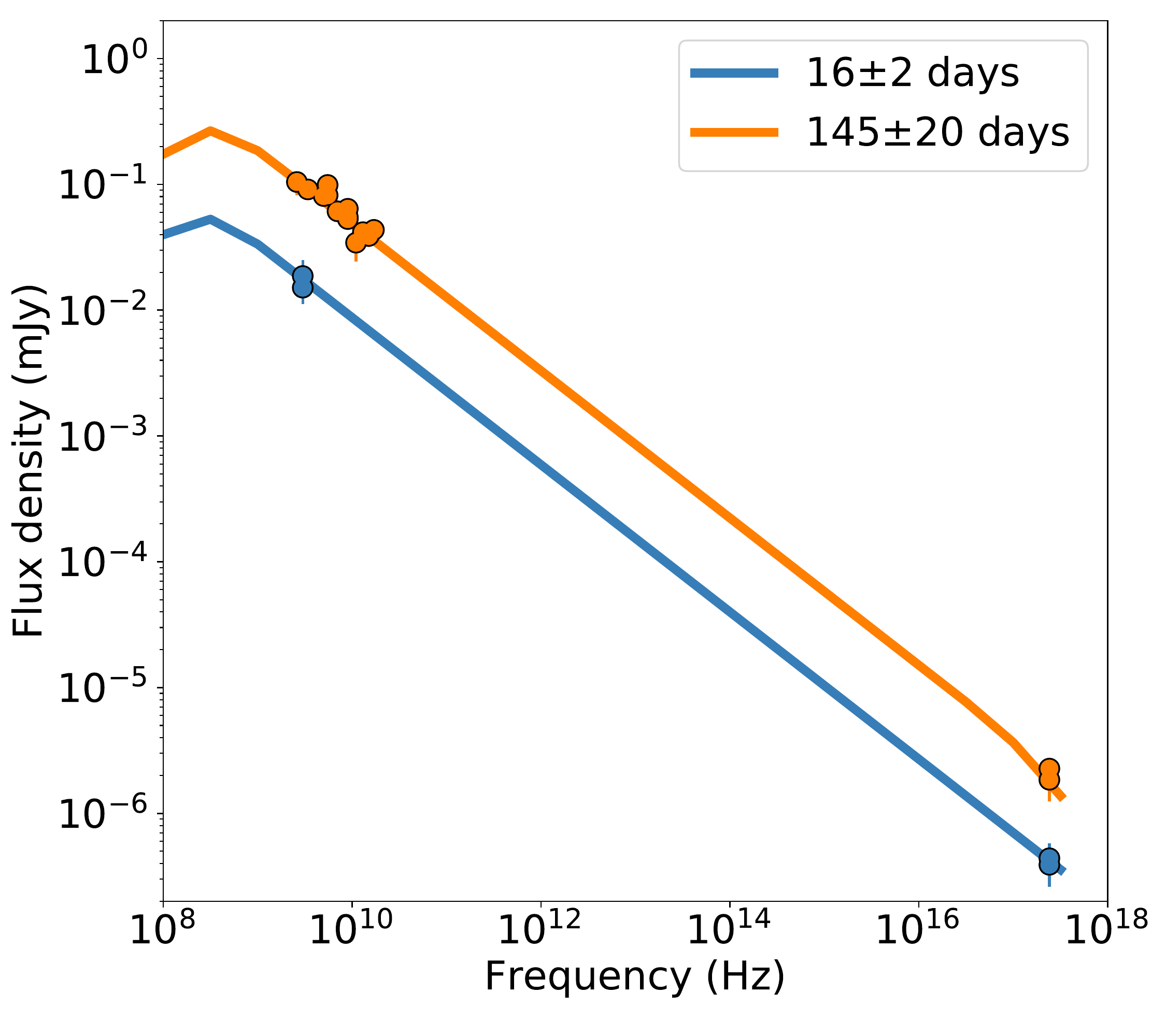}
 \caption{Broadband spectra of the transient produced by the external
   shock from a structured jet seen off-axis. The afterglow spectrum
   from the radio to the X-rays is shown at two epochs. In blue, the
   time of the earliest radio and X-ray measurements ($\sim$16 days after
   the trigger), in orange, the epoch at which the
   radio light curve will peaks, approximately 145 days after the
   trigger. The displayed data have been taken within a few days
   of the time at which the spectra were computed (as specified in the
   legend).
   \label{fig:spex}}
 \end{figure}

 \begin{acknowledgments} We thank J. Iván Castorena for help with
   gathering observational data from the literature. DL acknowledges
   support from NASA ATP grant NNX17AK42G. RP acknowledges support by
   NSF award AST-1616157. The Flatiron Institute is supported by the
   Simons Foundation.
\end{acknowledgments}

\bibliography{prl}

\end{document}